\pdfoutput=1

\documentclass[aps,letterpaper,twocolumn,10pt,aps,notitlepage,nofootinbib,nobibnotes]{revtex4-1}

\usepackage{amsfonts}
\usepackage{amsmath}
\usepackage{bm,graphicx}
\usepackage{hyperref}
\usepackage{epsfig}
\usepackage{verbatim}   
\usepackage[subfigure]{graphfig}
\usepackage{epstopdf}
\usepackage{dcolumn}
\usepackage{color}
\usepackage{tabularx}
\usepackage{braket}
\usepackage{float}
\usepackage{tikz}
\usepackage[percent]{overpic}
\graphicspath{{./figures2/}}
\usepackage{subfigure}
\usepackage{mwe}
\usepackage{soul}
\usepackage{ulem}

\usepackage[scale=0.8,bmargin=3cm,footnotesep=2cm]{geometry}
\usepackage[flushmargin]{footmisc}

\hypersetup{colorlinks, linkcolor = [rgb]{0,0.0,0.75}, citecolor = [rgb]{0,0.0,0.75}, urlcolor = [rgb]{0,0.0,0.75}}

\usepackage[utf8]{inputenc}
\usepackage{lineno}
\def\be{\begin{equation}}
\def\ee{\end{equation}}
\def\bea{\begin{eqnarray}}
\def\eea{\end{eqnarray}}

\newcommand{\nn}{\nonumber}
\newcommand{\la}{\label}

\definecolor{green}{rgb}{0,.5,0}

\begin{document}

\title{Neutral Weak Form Factors of Proton and Neutron}

\author{Raza Sabbir Sufian}
\vspace*{-0.5cm}

\affiliation{
\mbox{Department of Physics and Astronomy, University of Kentucky, Lexington, KY 40508, USA}}
\email{sufian@jlab.org}

\begin{abstract}
We determine the nucleon neutral weak electromagnetic form factors $G^{Z,p(n)}_{E,M}$ by combining results from light-front holographic QCD and lattice QCD calculations. We deduce nucleon electromagnetic form factors from light-front holographic QCD which provides a good parametrization of the experimental data of the nucleon electromagnetic form factors in the entire momentum transfer range and isolate the strange quark electromagnetic form factors $G^{s}_{E,M}$ using lattice QCD. From these calculations, we obtain precise estimates of the neutral weak form factors in the momentum transfer range of $0\,\text{GeV}^2\leq Q^2 \leq 0.5 \,\text{GeV}^2 $. From the lattice QCD calculation, we present $Q^2$-dependence of the strange quark form factors. We also deduce the neutral weak Dirac and Pauli form factors $F_{1,2}^{Z,p(n)}$ of the proton and the neutron. 
\end{abstract}


\maketitle

\section{Introduction}

In the electron elastic scattering from a hadron, parity-violating asymmetry arises from the interference of weak and electromagnetic amplitudes where the neutral weak current scattering is mediated by the $Z$-boson exchange. Because the weak current contains both vector and axial vector contributions, it violates parity and this property of the neutral weak current has been the main interest of the parity-violating (PV) experiments~\cite{Mueller:1997mt,Spayde:2003nr, Beise:2004py,
Aniol:2004hp, Maas:2004ta, Maas:2004dh, Aniol:2005zf, Aniol:2005zg,
Armstrong:2005hs, Acha:2006my, Androic:2009aa, Baunack:2009gy, Ahmed:2011vp}. These PV experiments are important as they allow measurements of the standard model parameters related to $Z$-boson couplings and search for new PV interactions beyond the standard model. When electroweak (EW) radiative corrections~\cite{Marciano:1990dp,Musolf:1993tb} are taken into account, the neutral weak electric and magnetic form factors $G^{Z,p}_{E,M}$ of the nucleon, under the assumption of isospin symmetry, can be expressed in terms of nucleon electric ($G^{\gamma,p(n)}_E$) and magnetic ($G^{\gamma,p(n)}_M$) form factors and a contribution from the strange (s) quarks as~\cite{Kaplan:1988ku,Mckeown:1989ir,Beck:1989tg,Beise:2004py},
\bea \la{eq1}
&&G^{Z,p(n)}_{E,M}(Q^2)= \frac{1}{4}\bigg[(1\!-\!4\sin^2 \theta_W)(1\!+\!R^{p(n)}_V) G^{\gamma , p(n)}_{E,M}(Q^2)\nn \\
 &&-(1\!+\!R^{n(p)}_V)G^{\gamma , n(p)}_{E,M}(Q^2)\! -\!(1\!+\!R_V^{(0)})G^{s}_{E,M}(Q^2)\bigg],
\eea
where the subscript $E(M)$ stands for the electric(magnetic) form factor (FF) and the superscript $p(n)$ stands for the proton(neutron). Under the isospin symmetry, the strange electromagnetic form factor (EMFF) is the same for the proton and neutron, i.e. $G^{s,p}_{E,M}=G^{s,n}_{E,M}=G^{s}_{E,M}$. $R^{p(n)}_V$ and $R^{(0)}_V$ are radiative corrections to the vector form factors calculated in Ref.~\cite{Marciano:1990dp} and translated into the $\overline{\text{MS}}$-scheme in Ref.~\cite{Musolf:1993tb}. The updated analysis of these radiative corrections can be found in Ref.~\cite{Erler:2004in} and we use the values listed in Ref.~\cite{Liu:2007yi} for the subsequent calculations. \\ 

The first measurement of the proton neutral weak magnetic form factor $G^{Z,p}_{M}$ from PV asymmetry in the polarized $\vec{e}-p$ scattering experiment was performed by the SAMPLE collaboration. Performed at a momentum transfer of {\mbox{$Q^2=0.1\,\text{GeV}^2$,}} the neutral weak magnetic form factor was found to be $G^{Z,p}_{M}(Q^2=0.1 \,\text{GeV}^2) = 0.34(11)\, \text{nucleon magneton (n.m.)}$ which corresponds to a value of $G^s_M(Q^2=0.1\,\text{GeV}^2)=0.23(44)$ n.m~\cite{Mueller:1997mt}. In an updated analysis Ref.~\cite{Forest:1998zr} of the SAMPLE data, one of the authors from Ref.~\cite{Mueller:1997mt} obtained PV asymmetry $A = (-5.22\pm2.24\pm0.62)\times10^{-6}$ compared to the $A=(-6.34\pm1.45\pm0.53)\times10^{-6}$ at $Q^2 = 0.1 \text{GeV}^2$ reported in Ref.~\cite{Mueller:1997mt}. Both of these PV asymmetries agree within uncertainties. While extracting $G^{Z,p}_M$ using Eq.~(\ref{eq1}), the author in Ref.~\cite{Forest:1998zr} used radiative corrections from Ref.~\cite{Musolf:1993tb} instead of the radiative corrections~\cite{Musolf:1990ts} that were used in Ref.~\cite{Mueller:1997mt}. The author in Ref.~~\cite{Forest:1998zr} obtained $G^{Z,p}_{M}(Q^2=0.1 \text{GeV}^2) = 0.29(16)\, \text{n.m.}$ which corresponds to $G^s_M(Q^2=0.1\,\text{GeV}^2)=0.49(65)$ n.m.. More technical details of this updated analysis, such as the inclusion of shutter closed asymmetries in the experiment, scintillation measurements, etc. are beyond the scope of this work and interested readers are referred to Ref.~\cite{Forest:1998zr} for more discussion.  Another reanalysis~\cite{Spayde:2003nr} of the SAMPLE data with three major modifications implemented, such as a developed Monte-Carlo simulation of the full experimental geometry, consideration of background associated with the threshold photo-pion production which was not included in Ref.~\cite{Mueller:1997mt}, and a different way of analyzing background coming from charged particles resulted in a measured asymmetry $A=(-5.61\pm0.67\pm0.88)\times10^{-6}$ which corresponds to $G^s_M(Q^2 = 0.1$ GeV$^2$) = 0.37(33) n.m.. A large positive value of $G^s_M$ corresponds to a $G^{Z,p}_M<0.40$ n.m. at $Q^2=0.1\,\text{GeV}^2$. Recent lattice QCD calculations favor a negative and small value of $G^s_M(0)$~\cite{Green:2015wqa,Sufian:2016pex}. To date, no individual experiment provides high precision measurements of the nucleon neutral weak FFs in a wide range of $Q^2$. By considering the weak axial vector form factor $G^e_A$ as an input, it is possible to separate the Sachs electric and magnetic FFs by combining PV asymmetry measurements from the experimental data. However, because of the complexity of the experiments, rather sizable uncertainties in the value of $G^e_A$ and the lack of knowledge of its $Q^2$ behavior, the extracted value of nucleon strange EMFFs from PV-asymmetry data vary widely in different experiments and global fits~\cite{Liu:2007yi,Gonzalez-Jimenez:2014bia,Young:2006jc}. Although the typical EW radiative corrections are expected to be $\mathcal{O}(\alpha)$, the tree-level suppression of the interaction in the PV $\vec{e}-p$ scattering makes the radiative corrections to $G^e_A$ more significant and radiative corrections involving the strong interaction are not clearly known~\cite{Zhu:2000gn}, extraction of $G^{Z,p(n)}_{E,M}$ from the PV scattering experiments is a tremendous challenge. One anticipates that with a reliable first-principles estimate of $G^s_{E,M}$, one can also give a prediction to the neutral weak FFs of the proton and the neutron without a prior knowledge of $G^e_A(Q^2)$ according to Eq.~(\ref{eq1}).\\

 The main goal of this article is to calculate the neutral weak FFs of the proton and the neutron $G^{Z,p(n)}_{E,M}$, $F_{1,2}^{Z,p(n)}$ by combining results of the strange quark EMFFs from the lattice QCD calculation in Ref.~\cite{Sufian:2016pex} and nucleon EMFFs calculated from the light-front holographic QCD in Ref.~\cite{Sabbir}. From the lattice QCD calculation, we isolate the $s$-quark contributions to the nucleon EMFFs and obtain the $Q^2$-dependence of the $s$-quark Sachs electric and magnetic FFs in the momentum transfer range of $0\leq Q^2\leq 0.5\,\text{GeV}^2$. In principle, one can use a parametrization of the strange quark EMFF to calculate $G^{s}_{E,M}(Q^2)$ in Eq.~(\ref{eq1}), for example, a parametrization given in Ref.~\cite{Hemmert:1998pi}. However, one has to have a prior knowledge of the strange quark magnetic moment $G^s_M(0)$ and the $Q^2$ behavior of $G^{s}_{E,M}(Q^2)$ in the nonperturbative region, which with a proper estimate of uncertainties, are not accurately known from phenomenological models at this moment. Also, three different global analyses of the experimental data give $G^s_{M}(Q^2= 0.1\, \text{GeV}^2)$ consistent with zero within their uncertainties and differ in sign in their central values~\cite{Liu:2007yi,Gonzalez-Jimenez:2014bia,Young:2006jc}. Therefore, we only use the  $G^{s}_{E,M}(Q^2)$ determined from the first-principles lattice QCD calculation in a momentum transfer range of $0\leq Q^2\leq 0.5$ GeV$^2$ where the statistical and systematic uncertainties can be estimated in a reliable way.\\
 
It has been shown in Ref.~\cite{Sabbir} and in Sec.~\ref{LFEMFF} that light-front holographic QCD (LFHQCD) can describe an extensive set of experimental data of the nucleon EMFFs in any momentum transfer range with high precision. The higher Fock-states probabilities in the following LFHQCD calculation of the nucleon EMFFs are obtained by fitting the experimental data. One can alternatively use the experimental data summarized in Refs.~\cite{Qattan:2017lhm,Punjabi:2015bba} or parametrization to the experimental data, such as Kelly's parametrization~\cite{Kelly:2004hm} of the nucleon FFs, or lattice QCD calculations of the nucleon EMFFs. Neutron FF measurements are challenging and the experimental data are not still up to the desired level of precision compared to the experimental measurements of proton EMFFs. On the other hand, lattice QCD calculations of the neutron EMFF is also very challenging, especially the neutron Sachs electric FF. From lattice QCD simulation at the physical pion mass, in Refs.~\cite{Alexandrou:2016hiy,Alexandrou:2017ypw}, it is seen that neutron Sachs EFF is particularly noisier than the other Sachs form factors and undershoots the experimental data points in the momentum transfer region of $Q^2<0.5$ GeV$^2$. Tremendous improvements have been achieved in calculating nucleon EMFF from lattice QCD calculations over the past years and more statistics is required to reproduce the experimental data with controlled systematics. A development toward such a calculation using physical pion mass and several lattice volumes is underway and needs more computer resources at this stage to include both the valence and disconnected light-sea quarks contribution to the nucleon EMFFs calculations. Since the LFHQCD predictions of nucleon EMFFs describe the experimental data very well, at this stage, instead of using experimental data of the nucleon EMFFs or lattice QCD calculation of nucleon total EMFFs, we use LFHQCD formalism to calculate nucleon EMFFs in Eq.~(\ref{eq1}), which, in a sense is just a parametrization of the average of the world experimental data. As we will discuss below, the nucleon EMFFs calculated from LFHQCD has a model uncertainty of about 10\% based on the value of the emerging confinement scale and systematic uncertainties associated with the free parameters used in the calculation. While extracting neutral weak FFs of the neutron, along with different sources of model uncertainties associated with this LFHQCD calculation and the statistical and systematic uncertainties of the lattice QCD calculation of $G^s_{E,M}(Q^2)$, we also include the uncertainty coming from SU(6) symmetry breaking associated with the free parameter $r$ used to calculate neutron electric FF in the following calculation.\\
 
Light-front holographic QCD developed in Refs.~\cite{Brodsky:2006uqa,deTeramond:2008ht,deTeramond:2013it} provides new insights into the quantitative determination of hadron mass spectra and FFs within a relativistic frame-independent first-approximation to the light-front QCD Hamiltonian. This new approach to hadronic physics follows from an approximate mapping of the Hamiltonian equations in the Anti-de Sitter (AdS) space to the relativistic semiclassical bound-state equations in the light front~\cite{deTeramond:2008ht,deTeramond:2013it}. This connection gives an exact relation between the holographic variable $z$ of the AdS space and the invariant impact light-front variable $\zeta$ in the physical space-time~\cite{Brodsky:2006uqa,deTeramond:2008ht}. The LFHQCD approach incorporates superconformal quantum mechanics and captures the relevant aspects of color confinement based on a universal emerging single mass scale $\kappa = \sqrt{\lambda}$~\cite{Brodsky:2013ar, deTeramond:2014asa,Dosch:2015nwa,Brodsky:2016yod,deTeramond:2016bre,Brodsky:2010ur,Deur:2016tte}. In the LFHQCD approach baryons correspond to $N_c=3$~\cite{Brodsky:2014yha}. Nucleon FFs determined within this nonperturbative framework incorporate vector dominance~\cite{Sakurai:1960ju} at small $Q^2$ and correct leading twist-$\tau$ scaling or power law fall-off for hard scattering independent of the specific dynamics at large $Q^2$~\cite{Brodsky:1973kr,Matveev:1973ra}. The most recent analysis of the nucleon EMFFs and their flavor-decomposition in the spacelike region from LFHQCD shows remarkable agreement with the experimental data when effects of the pion cloud and SU(6) spin-flavor symmetry breaking for the neutron are considered~\cite{Sabbir}. With the confinement scale fixed by hadron spectroscopy and the anomalous magnetic moments of proton and neutron fixed by experiment, only three additional free parameters are necessary to describe an extensive set of data of the nucleon EMFFs. It is important to note that, a central goal of hadron physics is to not only successfully predict
these dynamical observables but to also accurately account for the spectroscopy of hadrons. This new approach to hadron physics predicts universal linear Regge trajectories and slopes
in both orbital angular momentum and radial excitation quantum numbers, the appearance of
a massless pion in the limit of zero-mass quarks, and gives remarkable connections between the
light meson and nucleon spectra~\cite{Dosch:2015nwa,deTeramond:2014asa,Brodsky:2016yod}.\\ \\
Conventionally, we omit the unit nucleon magneton (n.m.) for the form factors in the rest of the paper. We also use the simple notations $G^{\gamma, p(n)}_{E,M}\equiv G^{p(n)}_{E,M}$ and $F_{1,2}^{p(n)}$ to describe the parity-conserving nucleon EMFFs in the following calculations and figures. 
\section{Nucleon electromagnetic form factors in light-front holographic QCD} \la{LFEMFF}
We now present calculation of the nucleon EMFFs in the framework of light-front holographic QCD. The details of the calculation can be found in our recent work~\cite{Sabbir}. Considering pion cloud effect and breaking of SU(6) spin-flavor symmetry for the neutron Dirac FF, we write proton and neutron Dirac and Pauli FFs in terms of a different combination of twist operators following Ref.~\cite{Sabbir} as, 
\bea \label{protonF1} 
F_1^p(Q^2) &=& F_{\tau=3}(Q^2)  \\  \la{protonF2}
F_2^p(Q^2) &=& \chi_p [(1-\gamma_p)F_{\tau=4}(Q^2) + \gamma_p F_{\tau=6}(Q^2)],
\eea
for the proton, with $\chi_p=1.793$ the proton anomalous moment, and 
\bea \la{neutronF1}
F_1^n(Q^2) &=&  -\frac{1}{3}r\left[F_{\tau=3}(Q^2) - F_{\tau=4}(Q^2) \right] ,  \\ \la{neutronF2} 
F_2^n(Q^2) &=& \chi_n \!\left[(1- \!\gamma_n)F_{\tau=4}(Q^2) + \gamma_n F_{\tau=6}(Q^2) \right],
\eea
for the neutron, with $\chi_n=-1.913$ and $r=2.08$ a free-parameter required to properly match to the experimental data as discussed in~\cite{Sabbir}. $\gamma_p$ and $\gamma_n$ in Eqs.~(\ref{protonF2}) and (\ref{neutronF2}) are the probabilities associated with the inclusion of the higher Fock components $\ket{qqqq\bar{q}}$ in the proton and neutron spin-flip EM transition amplitude, respectively. This additional $\ket{q\bar{q}}$ contribution to the nucleon wave function from higher Fock components is relevant at larger distances and is usually interpreted as a pion cloud. The twist-$\tau$ of a particle is defined here as the power behavior of its light-front wave function near $\zeta =0$: $\Phi  \sim \zeta^\tau$. For ground state hadrons the leading twist is the number of constituents. When computing nucleon FFs one has to constrain the asymptotic behavior of the leading fall-off of the FFs to match the twist of the nucleon's interpolating operator, {\it i.e.} $\tau = 3$, to represent the fact that at high virtualities the nucleon is essentially a system of 3 weakly interacting quarks. For a given twist $\tau$, the FFs on the right-hand side in Eqs.~(\ref{protonF1})-(\ref{neutronF2}) can be written by shifting the vector meson poles to their physical locations as
\bea \label{FFtau} 
  F_{\tau}(Q^2) =  \frac{1}{{\left(1 \!+\! \frac{Q^2}{M^2_{\rho_{n=0}}} \right)}\!
 \left(1\! +\! \frac{Q^2}{M^2_{\rho_{n=1}} } \right)\! \cdots  \left(1\! +\! \frac{Q^2}{M^2_{\rho_{n = \tau -2}} } \right)},\nn \\
 \eea
 where  
\bea \la{eq6}
-Q^2= M^2_{\rho_n} = 4 \kappa^2 \left(n+ \frac{1}{2} \right), \,\, n=0,1,2, ... \quad .
\eea
This shift of the poles of the conserved AdS current form is completely ad-hoc and motivated by adjusting to the observed poles of the EM current in the strong sector. The ground-state mass of the rho($\rho$) meson, $M_{\rho_{n=0}}\equiv M_\rho = 0.775 \,\text{GeV}$ gives the value of $\kappa =   M_\rho / \sqrt 2 = 0.548\, \text{GeV}$, where $\kappa = \sqrt{\lambda}$ is the emerging confinement scale~\cite{Brodsky:2013ar}.
 Eq.~(\ref{FFtau}) is expressed as a product of $\tau -1$ poles along the vector meson Regge radial trajectory in terms of the $\rho$ vector meson mass $M_\rho$ and its radial excitations. 
The expression for the FF (\ref{FFtau}) contains a cluster decomposition: the hadronic FF factorizes into the $i = N - 1$ product of twist-2 monopole FFs evaluated at different scales~\cite{deTeramond:2016pov} ($N$ is the total number of constituents of a given Fock state):
 \bea
F_{i }(Q^2) = F_{\tau = 2}\left(Q^2\right)  \,F_{\tau = 2} \left(\tfrac{1}{3}Q^2\right)\,  \cdots\,    F_{\tau = 2} \left(\tfrac{1}{ 2 \tau - 1} Q^2\right).\nn \\
\eea

The nucleon Sachs EMFFs are written as linear combinations of the Dirac and Pauli FFs as the following:
 \bea \la{SachsFF}
 G^{p(n)}_E(Q^2) &=& F_1^{p(n)}(Q^2) -\frac{Q^2}{4m_N^2} F_2^{p(n)}(Q^2), \nn \\
 G^{p(n)}_M(Q^2) &=& F_1^{p(n)}(Q^2) + F_2^{p(n)}(Q^2),
 \eea
These expressions of the Sachs electric and magnetic form factors will be used to obtain nucleon neutral weak Sachs EMFFs in Eq.~(\ref{eq1}). \\

One can now obtain the asymptotic values of $F_\tau$, $F^{p(n)}_{1,2}$ and $R_{p(n)} = \mu_{p(n)}G^{p(n)}_E/G^{p(n)}_M$ using Eqs.~(\ref{FFtau},\ref{eq6}) and the following results:
\bea \la{asymptotic}
\lim_{Q^2\to \infty}  \left(Q^2\right)^{\tau -1}F_\tau(Q^2)&=&M_{n=0}^2  \cdots  M^2_{n= \tau-2}\nn \\
&=&\kappa^{2 \tau -2} \prod_{n=0}^{\tau-2} (2 + 4n), \nn \\
\lim_{Q^2\to \infty}  Q^4
F^p_1(Q^2) &=& M_{n=0}^2 \,  M_{n=1}^2 = 12\,\kappa^4,\nn \\
\lim_{Q^2\to \infty}  Q^4
F^n_1(Q^2) &=& - \frac{1}{3} \, r \, M_{n=0}^2 \,  M_{n=1}^2 = - 4 r\,\kappa^4,\nn \\
\lim_{Q^2\to \infty} \!\!\! Q^6\!
F^N_2(Q^2)& =& \chi_N P^\gamma_{qqq / N} M_{n=0}^2 \,  M_{n=1}^2 \,  M_{n=2}^2\nn \\
&=& 120\,\chi_N P^\gamma_{qqq / N} \, \kappa^6,\nn \\
\lim_{Q^2\to \infty} 
R_p(Q^2) &=&  \!\mu_p\! \left(\!1\!-\! \frac{5}{2}(\mu_p\!-\!1) P^\gamma_{qqq/p} \, \frac{\kappa^2}{m_p^2}\right),\nn \\
\lim_{Q^2\to \infty} 
R_n(Q^2)&=& \mu_n \left(1+\frac{15 \, \mu_n P^\gamma_{qqq/n}}{2 r} \frac {\kappa^2}{m_n^2} \right),\nn \\
\eea
where $P^\gamma_{qqq / N} = (1 - \gamma_N)$, $N = p, n$, is the valence probability for the spin-flip EM transition amplitude. Possible logarithmic corrections are, of course, not predicted in this semiclassical model.
Keeping in mind that the gauge-gravity duality does not determine the spin-flavor structure of the nucleons, this one is conventionally included in the nucleon wave function using SU(6) spin-flavor symmetry. The departure of the free-parameter $r$ from unity may be interpreted as a SU(6) symmetry breaking effects in the neutron Dirac FF. Equations (\ref{protonF1}) and (\ref{neutronF1}) are the SU(6) results for the spin nonflip nucleon FFs in the valence configuration~\cite{Brodsky:2014yha,deTeramond:2012rt}. Equations~(\ref{protonF2}) and (\ref{neutronF2}) correspond to the extension of the phenomenological spin-flip nucleon FFs described in Refs.~\cite{Brodsky:2014yha,deTeramond:2012rt} and which incorporate the effect of twist-6 Fock components, i.e. the contribution of $\Ket{qqqq\bar{q}}$ components in the nucleon Pauli FFs. We obtain the probabilities $\gamma_p=0.27 \pm 0.03$ in Eq.~(\ref{protonF2}) and $\gamma_n = 0.38 \pm 0.05$ in Eq.~(\ref{neutronF2}) by fitting the world experimental data presented in the review article~\cite{Punjabi:2015bba} and the references therein. An attempt to include higher Fock component contribution in the proton or neutron Dirac FF results in a zero probability in the fits of the experimental data as discussed in Ref.~\cite{Sabbir}. The additional parameter $r$ accounts for the SU(6) symmetry breaking effects in the neutron and $r = 2.08\pm 0.09$ gives a proper match to the experimental data. In the fits of the experimental data of the nucleon EMFFs, we use $\gamma_p$, $\gamma_n$ and $r$ as free parameters by keeping $\kappa=0.548$ GeV fixed and obtain the values of the fit parameters with the quoted uncertainties. We make sure that the $\chi^2/d.o.f.$ for the fits of proton and neutron experimental data is in the vicinity of $1.0$. We include the uncertainties associated with these fit parameters and a model uncertainty associated with the value of $\kappa$, which we discuss in Sec.~\ref{MU}.

\subsection{LFHQCD Model Uncertainties} \la{MU}
LFHQCD, constrained by superconformal quantum mechanics~\cite{Dosch:2015nwa}, yields a semiclassical description of QCD that can be regarded as a first approximation to the full QCD. Therefore, for example, logarithmic terms due to quantum loops are absent in the model. This is reflected by the fact that the fitted values of the universal confinement scale $\kappa= \sqrt{\lambda}$ differ by about $10\%$ for different trajectories~\cite{Brodsky:2016yod}. We obtain from the $\rho$ -trajectory the value $\kappa = 0.537$ GeV, from the nucleon trajectory $\kappa=0.499$ GeV, and from a fit to the $\rho$ -mass alone $\kappa= M_\rho/\sqrt{2}=0.548$ GeV. Since the $\rho$ -pole is dominant for the nucleon FFs, we take this last value of $\kappa=0.548$ GeV as the default value in our calculation. However, for the low $Q^2$-region, the form of the nucleon wave function is important. Therefore we estimate the uncertainty in this region from the difference of the results obtained with the default value of $\kappa=0.548$ GeV and the result obtained with $\kappa =0.499$ GeV from the nucleon trajectory. We also consider the uncertainty of the nucleon FF by using the zero-probabilities ($\gamma_p=\gamma_n=0$) of the higher Fock components in the Pauli FFs when calculating neutral weak FFs. In the low-energy domain of our calculation, i.e. for $Q^2\leq 0.5\,\text{GeV}^2$, the largest uncertainty comes from the difference in $\kappa$-values and the uncertainty due to higher Fock components are very small. \\

\begin{figure}[htbp]
\centering
\subfigure[Nucleon Dirac FFs from LFHQCD multiplied by $Q^4$]{%
\includegraphics[height=5cm,width=8cm]{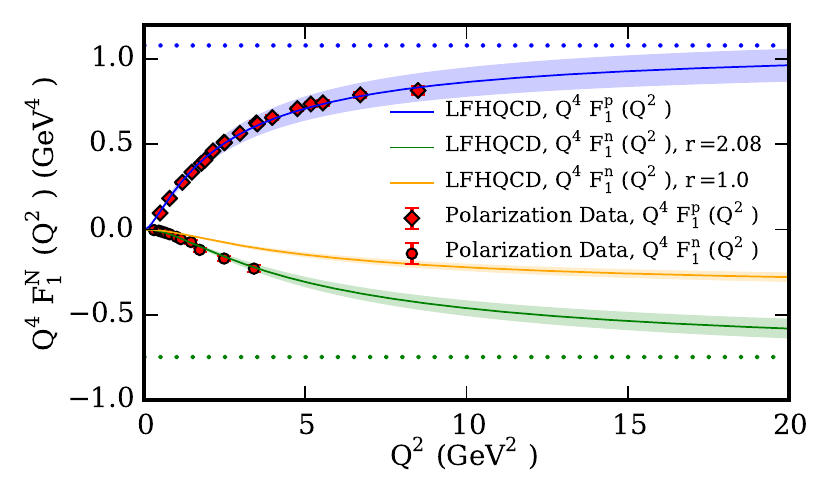}
\label{fig:subfigure1}}
\quad 
\subfigure[Nucleon Pauli FFs from LFHQCD multiplied by $Q^6$]{%
\includegraphics[height=5cm,width=8cm]{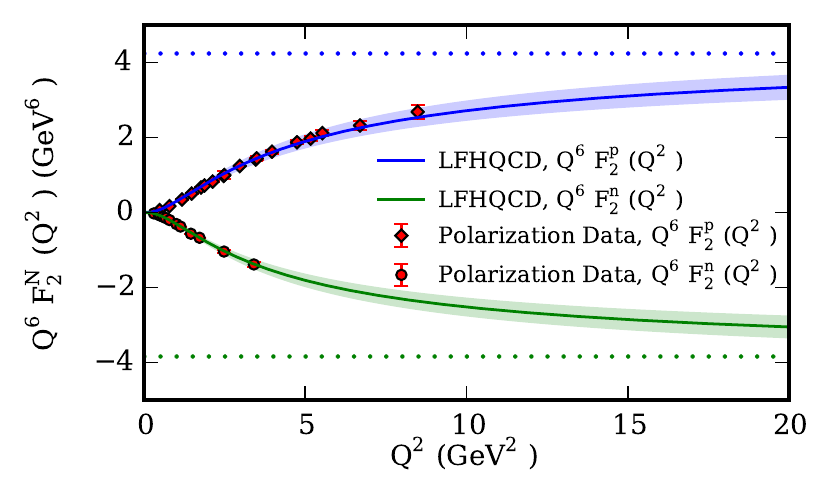}
\label{fig:subfigure2}}
\caption{Comparison of the LFHQCD results with selected world data~\cite{Riordan:2010id,Qattan:2012zf} for the Dirac and Pauli form factors for the proton and neutron. The orange line corresponds to the SU(6) symmetry limit for the neutron Dirac form factor. The dotted lines are the asymptotic predictions of the form factors from LFHQCD. The blue and green uncertainty bands are obtained from the variation of $\kappa$ determined by the nucleon and the $\rho$-trajectories.}
\label{fig:figureFF}
\end{figure}

We show in Fig.~\ref{fig:figureFF} and in Ref.~\cite{Sabbir} that nucleon EMFFs can be calculated very accurately using the above FF expressions (Eqs.~(\ref{protonF1})-(\ref{neutronF2})) and the available experimental data for both the proton and neutron at low and high $Q^2$ are described very well. The uncertainty bands in the Fig.~\ref{fig:figureFF} are obtained from the variation of $\kappa$ in the low and high~$Q^2$ domains. The asymptotic values of $Q^4F_1^{p(n)}$ and $Q^6F_2^{p(n)}$ are obtained from Eq.~(\ref{asymptotic}) for $\kappa=0.548$ GeV. Comparison of LFHQCD prediction of the nucleon FFs with other experimental data and flavor decomposition of the nucleon FFs can be found in Ref.~\cite{Sabbir}. \\

\begin{figure}[ht]
\includegraphics[height=5.5cm,width=8cm]{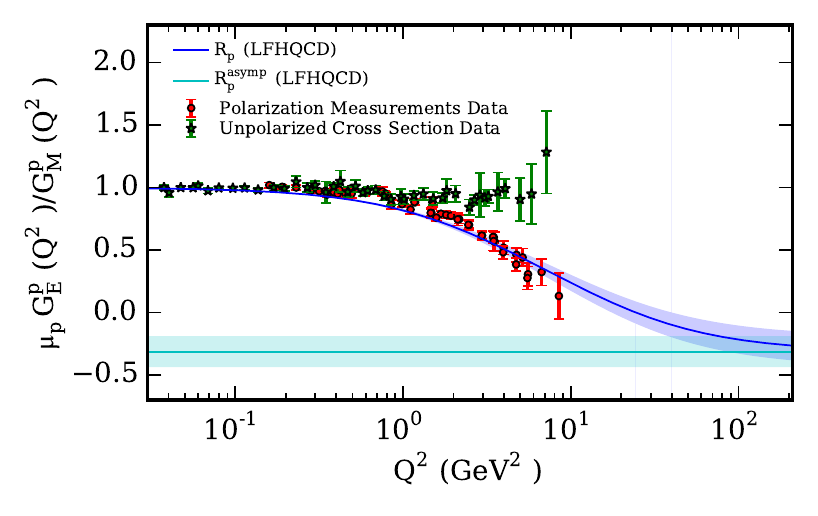}
\caption{\label{figRp} LFHQCD prediction and comparison with selected world data of the ratio $R_p = \mu_pG^{p}_E/G^{p}_M$ from unpolarized cross section measurements~\cite{Bartel:1973rf,Price:1971zk,Hanson:1973vf,Borkowski:1974mb} and polarization measurements~\cite{Jones:1999rz,Punjabi:2005wq,Gayou:2001qd,Puckett:2010ac,Jones:2006kf,Ron:2011rd,Crawford:2006rz}. The blue uncertainty band in the LFHQCD prediction of $R_p$ is obtained from the variation of $\kappa$ determined by the nucleon and the $\rho$-trajectories. The uncertainty in the cyan band of the asymptotic value $R_p^\text{asymp}(\infty)=-0.309$ is obtained form the difference between $\kappa=0.548$ GeV and $\kappa=0.537$ GeV.}
\end{figure}

It is shown in Fig.~\ref{figRp} that the uncertainty in the FFs for the variation in $\kappa$ does not diverge at very large $Q^2$, e.g. at $Q^2=200$ GeV$^2$, which is guaranteed by the fact that the hard-scattering power law fall-off is ensured for the FFs in LFHQCD formalism as mentioned earlier. We also show in Fig.~\ref{figRp} that the value of $R_p$ agrees with its asymptotic value and describes the experimental data in the entire $Q^2$ in a satisfactory way. Therefore, we use the LFHQCD formalism to calculate nucleon EMFFs in obtaining the nucleon neutral weak FFs in the rest of the calculation which basically gives a proper description to the average of the existing experimental data.

\section{Strange quark form factors from lattice QCD}\la{LQCD}
We have calculated the strange quark contribution to nucleon's magnetic moment and charge radius in Ref.~\cite{Sufian:2016pex} using the overlap fermion on the $(2+1)$ flavor RBC/UKQCD domain wall fermion (DWF) gauge configurations. Details of these ensembles are listed in Table~\ref{table1}. We use 24 valence quark masses in total for the 24I, 32I, 32ID, and 48I ensembles representing pion masses in the range $m_{\pi}\in$(135, 400) MeV to explore the quark-mass dependence of the $s$-quark FFs. 
\begin{table}
\begin{center}
\tabcolsep=0.03cm
\begin{tabular}{|c|c|c|c|c|c|}
\hline
Ensemble & $L^3\times T$  &$a$ (fm) & $m_s^{(s)}$(MeV) &  {$m_{\pi}$} (MeV)  & $N_\text{config} $ \\
\hline
24I~\cite{Aoki} & $24^3\times 64$& 0.1105(3) &120   &330  & 203    \\
\hline
32I~\cite{Aoki} &$32^3\times 64$& 0.0828(3) & 110   &300 & 309 \\
\hline
32ID~\cite{Blum} &$32^3\times 64$& 0.1431(7) & 89.4& 171 & 200\\
\hline
48I~\cite{Blum} &$48^3\times 96$& 0.1141(2) & 94.9   &139 & 81 \\
\hline
\end{tabular}
\caption{\label{table1} The parameters for the DWF configurations: spatial/temporal size, lattice spacing \cite{Aoki,Blum}, the sea strange quark mass under $\overline{\text{MS}}$ scheme at {2 GeV}, the pion mass corresponding to the degenerate light sea quark mass and the numbers of configurations used in this work.}
\end{center}
\end{table}
One can perform the model-independent $z-$expansion fit~\cite{Hill, Epstein}
 \bea \la{zexp}
G^{s,z-exp}(Q^2)\! =\! \sum^{k_{max}}_{k=0}\! a_k z^k, \, z=\!\frac{\sqrt{t_{\text{cut}}+Q^2}-\sqrt{t_{\text{cut}}}}{\sqrt{t_{\text{cut}}+Q^2}+\sqrt{t_{\text{cut}}}},\nn \\
\eea 
using the lattice data to extrapolate the $s$-quark magnetic moment and charge radius as shown in~\cite{Sufian:2016pex} and then use the fit parameters $a_k$ to interpolate $G^s_{E,M}$ values at various $Q^2$ for a given valence quark mass on the lattice. The available $Q^2$ on the 24I and 32I ensembles are $Q^2\in(0.22, 1.31)\,\text{GeV}^2$, on the 32ID ensemble are~$Q^2\in(0.07, 0.43)\,\text{GeV}^2$ and on the 48I ensemble are $Q^2\in(0.05, 0.31)\,\text{GeV}^2$. It is a common problem for lattice QCD calculations that the signal-to-noise-ratio decreases as one reaches the physical pion mass. From our study, we also find that the lattice results of $G^s_{E,M}(Q^2)$ near the physical pion mass $m_\pi=140\,\text{MeV}$ for the 48I ensemble~\cite{Blum} is noisier compared to the $G^s_{E,M}(Q^2)$ obtained from the lattice ensembles with heavier pion mass. Although the largest available momentum transfer we have on the 24I and 32I ensemble is $Q^2\sim 1.3$ GeV$^2$, the largest momentum transfer available on the 48I ensemble is $Q^2~\sim 0.31$ GeV$^2$. We note that the extrapolation of the nucleon strange EMFF starts to break down after $Q^2~\sim 0.4$ GeV$^2$ for the 48I ensemble and we therefore constrain the extrapolations of the 48I ensemble EMFF up to $Q^2=0.5$ GeV$^2$. It is important to note that the lattice QCD estimate of $G^s_{E,M}(Q^2)$ we present here is the most precise and accurate first-principles calculation of $s$-quark EMFFs to date. This is the only calculation at the physical pion mass where we have considered the quark mass dependence, with finite lattice spacing ($a$), volume corrections, and partial quenching effect to determine the $s$-quark EMFFs.\\ 

After the $Q^2$-interpolation, for a given $Q^2$ -value, we obtain 24 data points corresponding to different valence quark masses from 3 different lattice spacings and volumes and 4 sea quark masses including one at the physical point. We use chiral extrapolation formula following Ref.~\cite{Hemmert2} and volume correction following Ref.~\cite{Tiburzi}. The empirical formula for the global fit of the strange quark Sachs electric FF at a given $Q^2$ is
\bea \la{gsefit}
G^s_E (m_\pi,&&\!\!\!\!\! m_K,m_{\pi,vs}, a, L) = A_0 + A_1 m_K^2 +A_2 m_\pi^2 \nn\\
&+&A_3m_{\pi,vs}^2 + A_4 a^2 + A_5\sqrt{L} e^{-m_\pi L},
\eea
where $m_{\pi}/m_K$ is the valence pion/kaon mass and $m_{\pi, vs}$ is
the partially quenched pion mass {\mbox{$m_{\pi, vs}^2 = 1/2(m_{\pi}^2 + m_{\pi, ss}^2)$}} with $m_{\pi, ss}$ the pion mass corresponding to the sea quark mass. The $\chi^2/\text{d.o.f.}$ for different $Q^2$ global fit ranges between 0.7-1.13. For example, in the continuum limit, the global fit for $Q^2=0.25\,\text{GeV}^2$ results in the physical value of $G^s_E\vert_\text{phys}=0.0024(8)$, $A_1=0.58(30)$, $A_2=-0.29(15)$, $A_3=-0.003(9)$, $A_4=0.001(2)$, and $A_5=-0.001(3)$ with $\chi^2/\text{d.o.f.}=1.1$. One can consider the $\log(m_K)$-term in the chiral extrapolation of $G^s_E$ as shown in~\cite{Hemmert2}, however our analysis shows that this term does not have any effect on the global fit for our lattice data. A similar vanishing difference has been observed if one considers $e^{-m_\pi L }$ instead of $\sqrt{L}e^{-m_\pi L }$ term in the volume correction. For example, including the factor $\log(m_K)$ and $e^{-m_\pi L }$ instead of $\sqrt{L}e^{-m_\pi L }$ one obtains, $G^s_E\vert_\text{phys}=0.0026$ in comparison with $G^s_E\vert_\text{phys}=0.0024$ we get from~(\ref{gsefit}). We include these small effects in the systematics of the global fit results. We also consider a 20\% systematic uncertainty from the model-independent $z$-expansion interpolation coming from adding a higher order term $a_3$ while fitting the $G^s_E(Q^2)$ data. These uncertainties from the empirical fit formula and $z$-expansion are added to the systematics discussed in~\cite{Sufian:2016pex}. \\

\begin{figure}[ht]
\includegraphics[height=5cm,width=8cm]{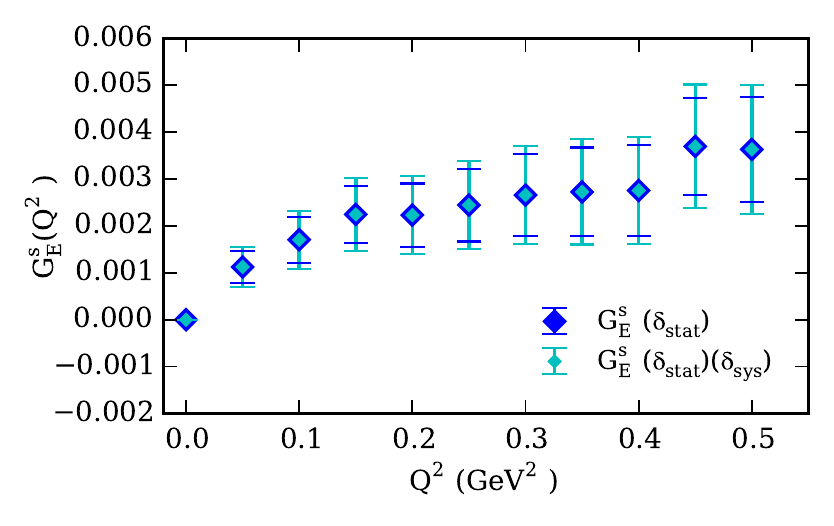}
\caption{\label{fig1} $Q^2$-dependence of the strange Sachs electric form factor. The blue error bars indicate the statistical uncertainties and the cyan error bars indicate the statistical and systematic uncertainties added in quadrature. }
\end{figure}

Figure~\ref{fig1} shows the $Q^2$-dependence of the $s$-quark Sachs electric form factor $G^s_E$ in the continuum limit, i.e. $m_\pi = m_{\pi, vs}\to 140\,\text{MeV}$, $a\to 0$, and $L\to \infty$ with the statistical and systematic uncertainties. The nonzero value of the strange Sachs electric form factor $G^s_E$ at any $Q\neq 0$ means that the spatial distribution of the $s$ and $\bar{s}$ quarks are not the same in the nucleon. If the distributions of the $s$ and $\bar{s}$ quarks were the same, their contribution to the nucleon electric FF would have the same magnitude with opposite signs. Since the net strangeness in the nucleon is zero, we have $G^s_E=0$ at $Q^2=0$.

\begin{figure}[ht]
\includegraphics[height=5cm,width=8cm]{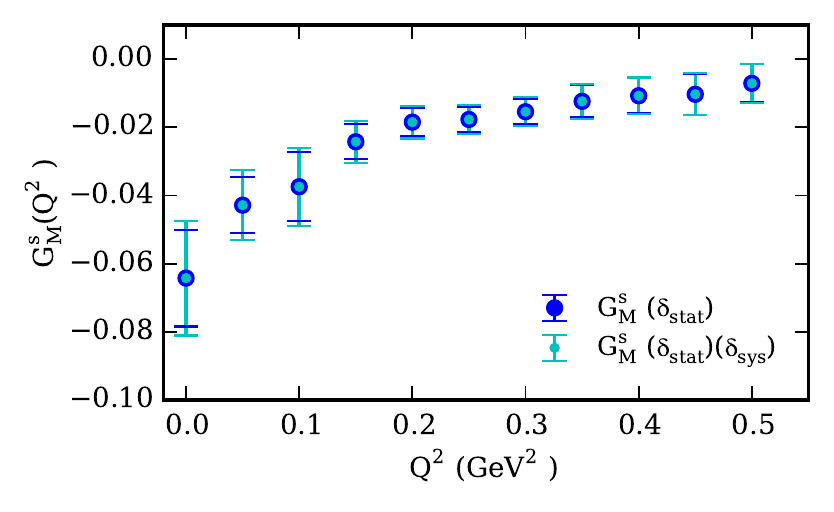}
\caption{\label{fig2} $Q^2$-dependence of the strange Sachs magnetic form factor. The blue error bars indicate the statistical uncertainties and the cyan error bars indicate the statistical and systematic uncertainties added in quadrature. }
\end{figure}

Similarly, we calculate the strange Sachs magnetic form factor $G^s_M$ at a particular $Q^2$ using the following global fit formula
 \bea \la{gsmfit}
G^s_M (&&\!\!\!\!\! m_\pi,m_K,m_{\pi,vs}, a , L) =A_0 + A_1 m_\pi + A_2 m_K \nn \\
&+& A_3m_{\pi,vs}^2 + A_4 a^2 + A_5m_\pi(1-\frac{2}{m_\pi L}) e^{-m_\pi L }, \nn \\
\eea
 where we have used a chiral extrapolation linear in $m_\pi$ and $m_{\text{loop}}=m_K$~\cite{Musolf, Hemmert1, Hemmert2, Chen}. For the volume correction we refer to Ref.~\cite{Beane}. From the global fit formula~(\ref{gsmfit}), for example, in the continuum limit at $Q^2=0.25\,\text{GeV}^2$, we obtain $G^s_M\vert_\text{phys}=-0.018(4)$, $A_1=0.04(3)$, $A_2 = -0.18(12)$, $A_3=-1.27(84)$, $A_4=0.008(6)$, and $A_5=0.04(5)$ with $\chi^2/\text{d.o.f.}=1.13$. From the fitted values of the parameters in the global fit formula~(\ref{gsmfit}), it is seen that the quark mass dependencies play an important role in calculating $G^s_M(Q^2)$ at the physical point. A 9\% systematic uncertainty from the model-independent $z-$expansion and an uncertainty from the empirical fit formula have been included as discussed in~\cite{Sufian:2016pex}. We obtain systematics from the global fit formula by replacing the volume correction by $e^{-m_\pi L }$ only and also by adding $m_{\pi,vs}$ term in the fit and include the difference in the systematics of the global fit results. The results of $G^s_{M}(Q^2)$ in the continuum limit are presented in Figure~\ref{fig2}. 
 
 \section{Calculation of neutral weak form factors}
 
\begin{figure}[ht]
\centering
\subfigure[Tree-level EM Feynman diagrams in elastic electron-nucleon scattering mediated by photon]{%
\includegraphics[height=2.0cm,width=3.2cm]{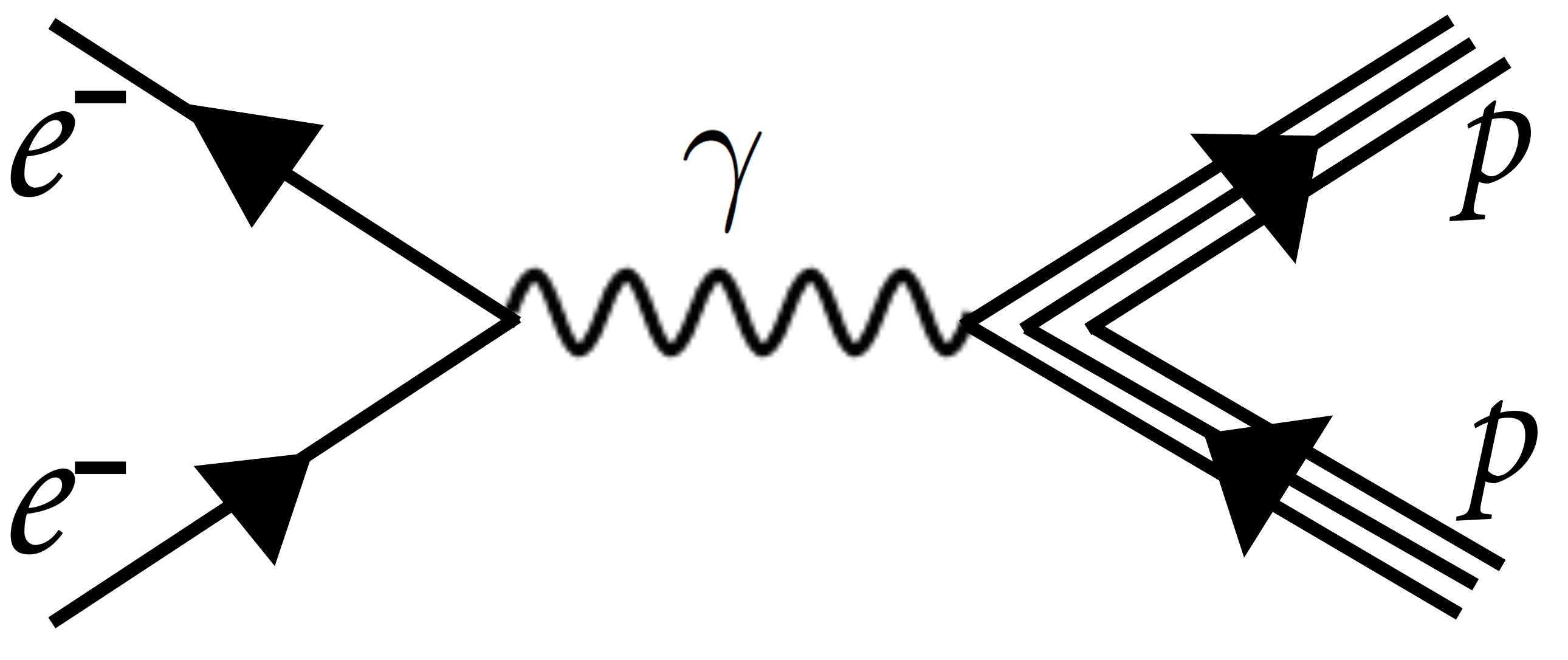}
\label{fig:subfigure1}}
\quad 
\subfigure[Tree-level weak Feynman diagrams in elastic electron-nucleon scattering mediated by neutral weak $Z$-boson]{%
\includegraphics[height=2.0cm,width=3.2cm]{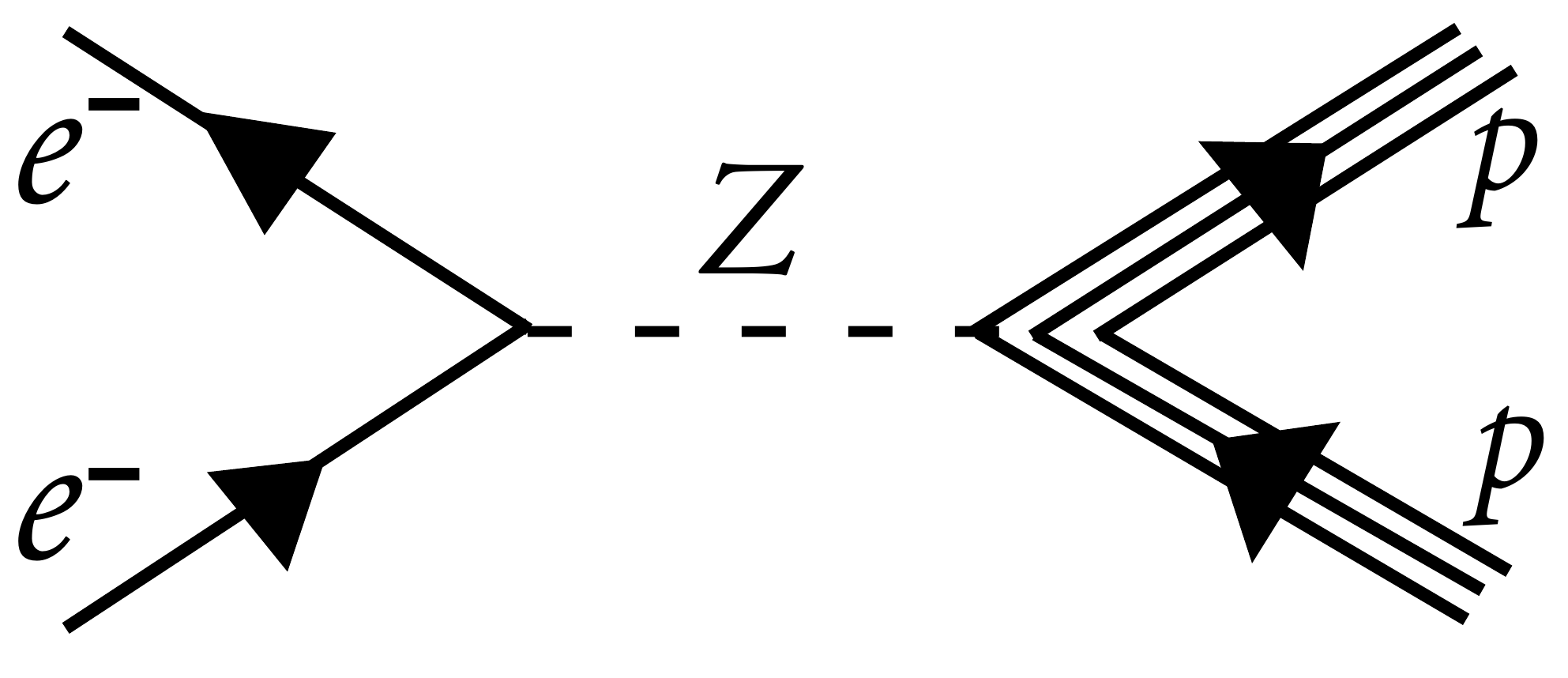}
\label{fig:subfigure2}}
\quad
\subfigure[Feynman diagram representing ``one-quark" radiative correction: vacuum polarization with leptons in the loop]{%
\includegraphics[height=2.0cm,width=3.2cm]{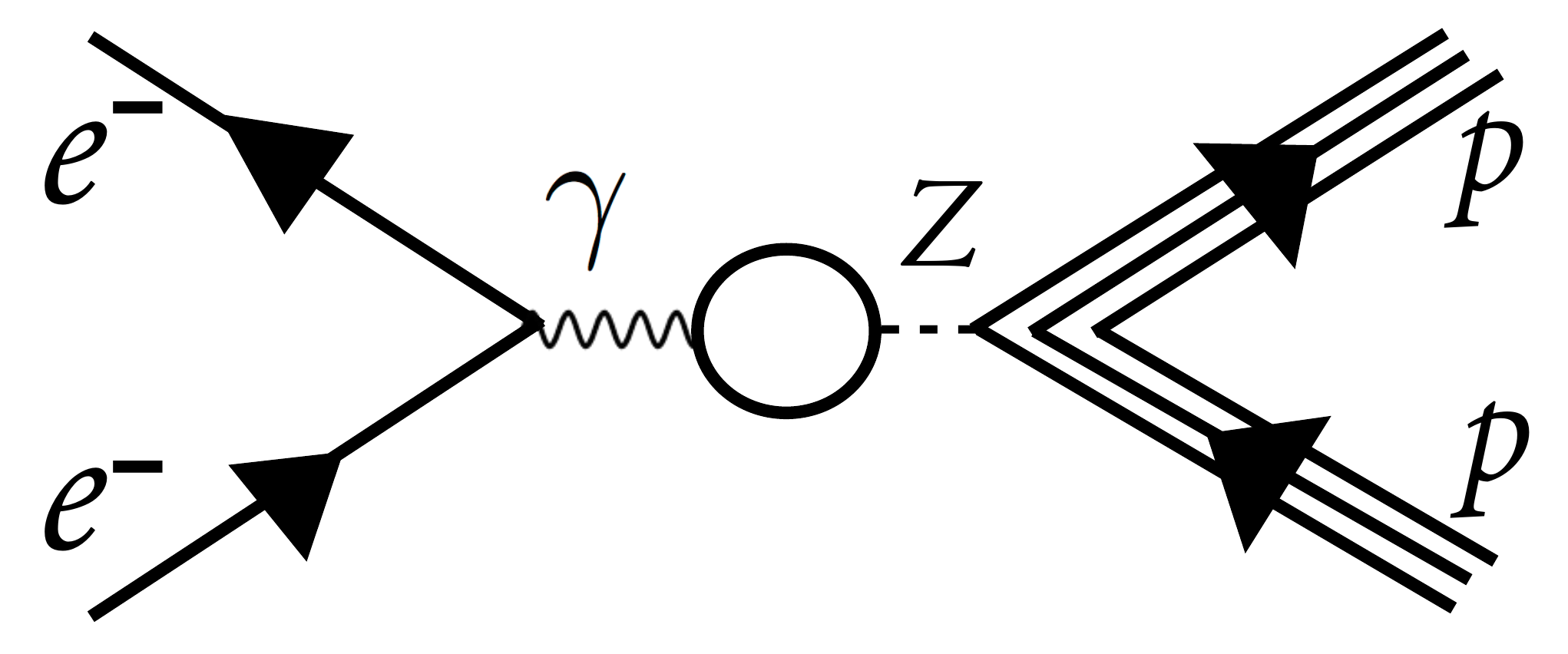}
\label{fig:subfigure4}}
\quad
\subfigure[Feynman diagram representing ``one-quark" radiative correction: $\gamma-Z$ box diagram ]{%
\includegraphics[height=2.0cm,width=3.2cm]{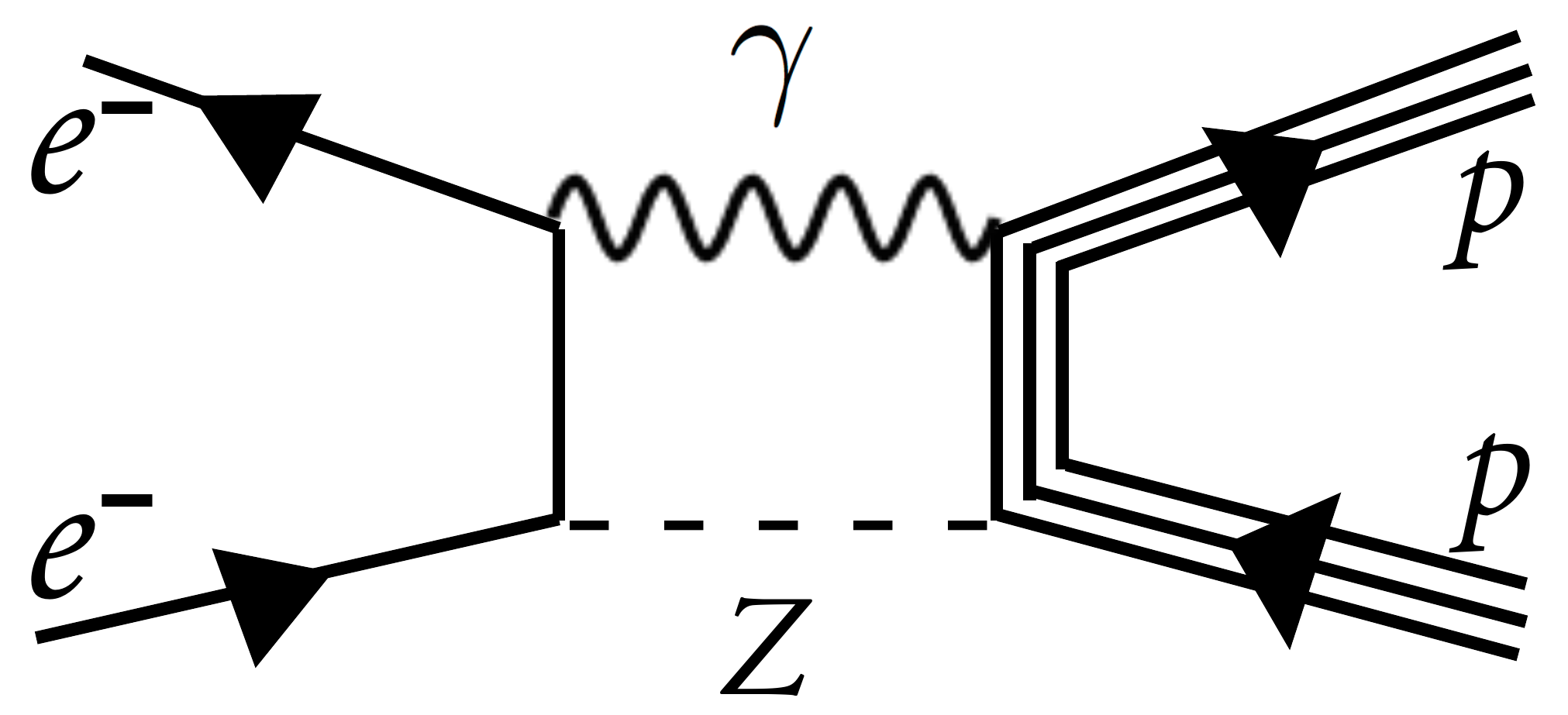}
\label{fig:subfigure3}}
\caption{Feynman diagrams representing tree-level EM and weak interactions and ``one-quark" radiative corrections.}
\label{fig:figure}
\end{figure}

\begin{figure}[htbp]
 \centering
\includegraphics[height=5cm,width=8cm]{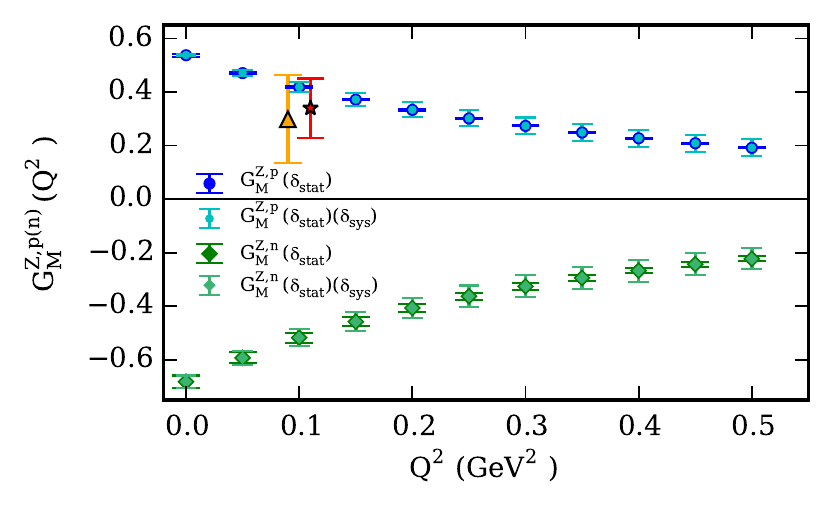}
\caption{$Q^2$-dependence of the proton and neutron neutral weak magnetic form factor $G^{Z,p(n)}_M$. The smaller uncertainties are from statistics alone of the lattice QCD calculation of $G^s_{E,M}(Q^2)$. The various systematic uncertainties from the LFHQCD model and lattice QCD calculation and the statistical uncertainties have been added in quadrature to obtain the final errors in the neutral weak FFs calculation. The red star is the experimental result from~\cite{Mueller:1997mt} and the orange triangle is from the analysis of SAMPLE proton data performed in~\cite{Forest:1998zr} at $Q^2=0.1\,\text{GeV}^2$ (with offset $Q^2$ for visibility).  \label{fig:3} }
\end{figure}

\begin{figure}[htbp]
 \centering
\includegraphics[height=5cm,width=8cm]{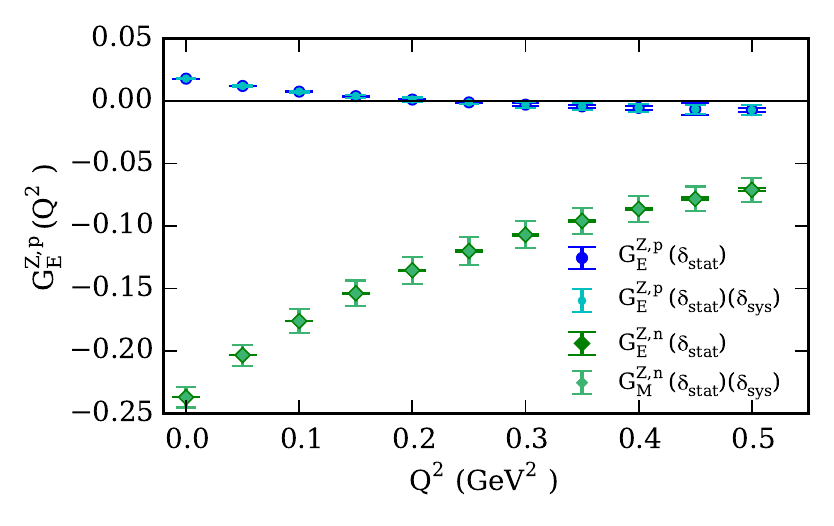}
\caption{$Q^2$-dependence of the proton and neutron neutral weak electric form factor $G^{Z,p(n)}_E$. The smaller uncertainties are from statistics of the lattice QCD calculation of $G^s_{E,M}(Q^2)$. The various systematic uncertainties from the LFHQCD model and lattice QCD calculation and the statistical uncertainties have been added in quadrature to obtain the final errors in the neutral weak FFs calculation.  \label{fig:4} }
\end{figure}

 Since the neutral weak $Z$-boson can have both vector and axial vector interactions, the amplitude of the $Z$-exchange can have both parity-conserving and parity-violating components. The parity-conserving and parity-violating $Z$-amplitudes in the electron-nucleon scattering can be written as
 \bea
 \mathcal{M}_Z^{PC} = \frac{G_F}{2\sqrt{2}}(g^i_Vl^\mu J^Z_{\mu }+g^i_A l^{\mu5}J^Z_{\mu5}) ,\\
  \mathcal{M}_Z^{PV} = \frac{G_F}{2\sqrt{2}}(g^i_Vl^\mu J^Z_{\mu 5}+g^i_A l^{\mu5}J^Z_\mu),
 \eea
where $G_F$ is the Fermi constant, $g^i_{V(A)}$ the weak vector(axial) charge of the fermions, $l^\mu(l^{\mu 5})$ the leptonic vector(axial) current, and $J^Z_{\mu}(J^Z_{\mu5})$ the nucleon vector(axial) current. In the electron-nucleon elastic scattering, the 
first-order interactions are mediated either by a photon($\gamma$) or a neutral weak $Z$-boson as shown in Figures~\ref{fig:subfigure1} and~\ref{fig:subfigure2}. The contributions to the weak FFs from additional diagrams in Figures~\ref{fig:subfigure4} and~\ref{fig:subfigure3}
should also be considered. Moreover, there can be contributions that involve strong interactions where $\gamma$ and $Z$-boson can interact with several quarks and these diagrams are not shown here. These ``many-quark" corrections are target specific and difficult to calculate; the calculations are model-dependent.
We use the LFHQCD predictions of nucleon electromagnetic form factors $ G^{p(n)}_{E,M}(Q^2) $ from Eq.~(\ref{SachsFF}) and $G^s_{E,M}(Q^2)$ from lattice QCD calculation in Eq.~(\ref{eq1}) to obtain the nucleon neutral weak FFs which are shown in Figures~\ref{fig:3} and~\ref{fig:4}. \\

We address several sources of systematic uncertainties coming from the LFHQCD model, such as from the variations in $\kappa$-value, from the higher Fock components probability parameters $\gamma_{p(n)}$ and from $r$ to estimate neutral weak FFs for the proton and neutron. When calculating the systematic uncertainties coming from the inclusion of higher Fock
components, we consider the difference between the FFs calculated with zero higher Fock components probability $\gamma_{p(n)}=0$ and probability $\gamma_{p(n)} = 0.27 (0.38)$ calculated by fitting the world average of the experimental data of nucleon EMFFs. Similarly, a systematic uncertainty in the neutron Dirac form factor is obtained by considering the SU(6) symmetry breaking parameter $r=2.08$ and neutron Dirac FF calculated without this free parameter, i.e. by considering $r=1$. Since, we are estimating neutral weak FFs in the $0\leq Q^2 \leq 0.5$ GeV$^2$, another systematic uncertainty comes from the difference of $\kappa=0.499$ GeV calculated from the nucleon trajectory and $\kappa=0.548$ GeV calculated from the $\rho$-mass. We also include systematic uncertainty from the systematics of the lattice QCD estimates of $G^s_{E,M}(Q^2)$ as discussed in Sec.~\ref{LQCD}. The statistical uncertainties in the neutral weak FFs come from the lattice QCD analysis of $G^s_{E,M}(Q^2)$. The total uncertainty of the neutral weak FFs at a specific $Q^2$-value is obtained by quadratically adding each source of systematic and statistical uncertainties and are shown separately from the statistical uncertainties in Figures~\ref{fig:3} and~\ref{fig:4}. The systematic uncertainties of LFHQCD model give the largest error in the estimates of neutral weak FFs.\\

As shown in Figure~\ref{fig:3}, our prediction of the proton neutral weak magnetic FF at $Q^2\approx0.1\text{GeV}^2$ is within the uncertainty of the experimental measurement by the SAMPLE collaboration $G^{Z,p}_{M}(0.1 \,\text{GeV}^2) = 0.34(11)$~\cite{Mueller:1997mt} but with better precision and the central value differs significantly. A model-dependent prediction of the proton neutral weak magnetic form factor can be found in Refs.~\cite{Silva:2001st, Silva:2002th}.\\

No experimental or theoretical estimates of the proton neutral weak electric FF and the neutron neutral weak electric and magnetic FFs have been reported in the literature to be compared with the calculated values in this work. While in the EM charge coupling, the electron couples to the proton and neutron with strengths 1 and 0 respectively, in the weak interaction the $Z$-coupling with the neutron is larger than the coupling to the proton. This can be seen from Figure~\ref{fig:4}: neutron neutral weak electric FF is much larger than the proton neutral weak electric FF. It can be seen in Figure~\ref{fig:4} that $G^{Z,p}_E$ becomes negative around $Q^2=0.25\,\text{GeV}^2$.

\begin{figure}[ht]
\centering
\subfigure[Neutral weak Dirac FF of the nulceon]{%
\includegraphics[height=5cm,width=8cm]{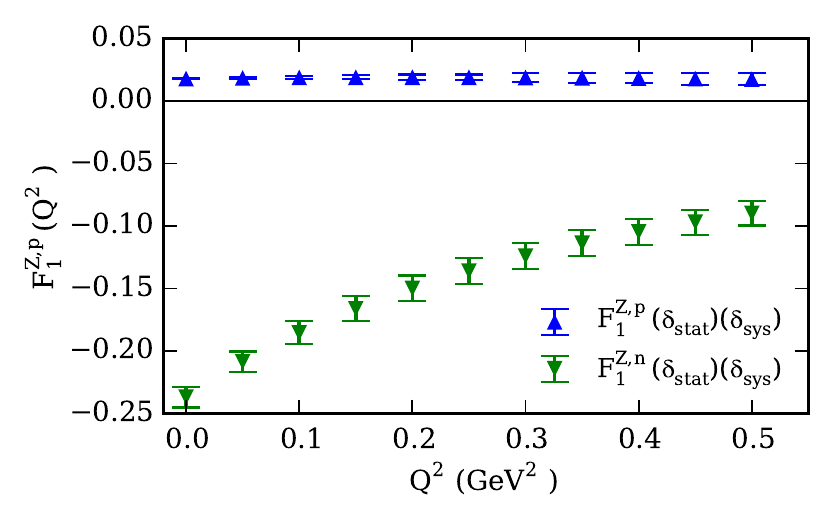}
\label{fig:subfiguref1z}}
\quad 
\subfigure[Neutral weak Dirac FF of the proton]{%
\includegraphics[height=5cm,width=8cm]{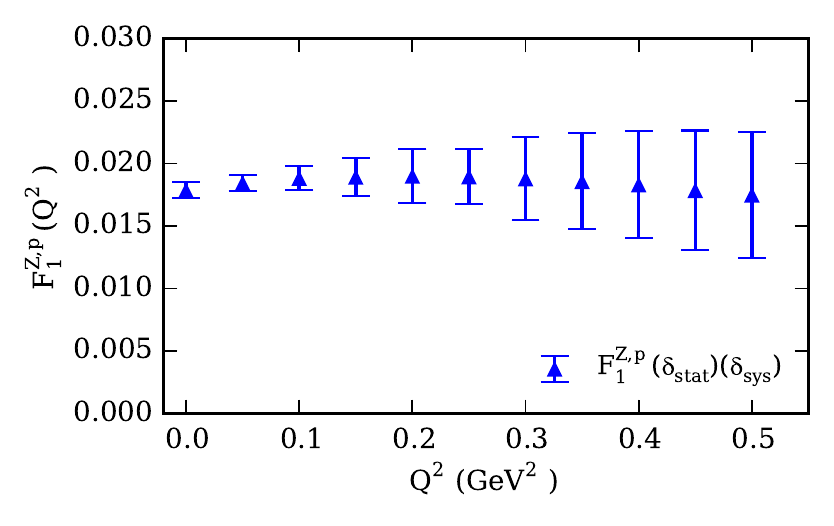}
\label{fig:subfiguref1zp}}
\caption{$Q^2$-dependence of the proton and neutron neutral weak Dirac form factor $F^{Z,p(n)}_1$. The $F^{Z,p}_1(Q^2)$ plot is shown separately in Figure~\ref{fig:subfiguref1zp} for better visibility in comparison with the $F^{Z,n}_1(Q^2)$ data. Statistical and systematic uncertainties are added in quadrature to obtain the final uncertainty. }
\label{fig:figureF1zpn}
\end{figure}

\begin{figure}[ht]
 \centering
\includegraphics[height=5cm,width=8cm]{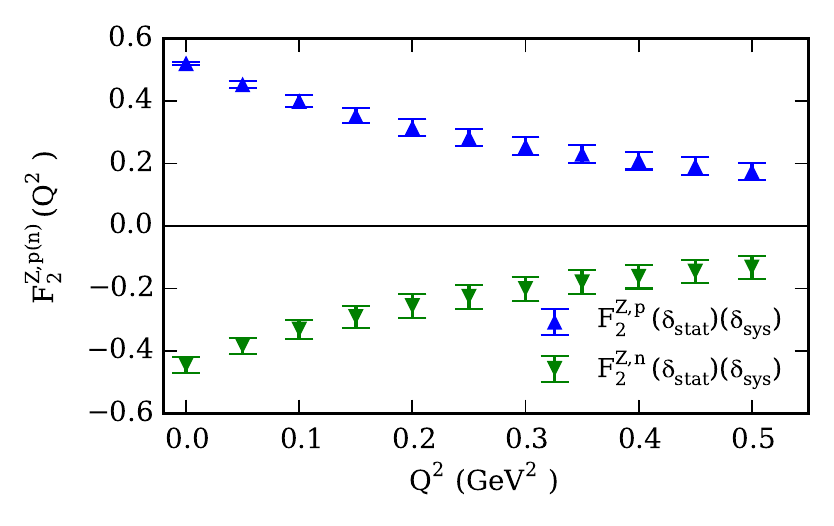}
\setlength\abovecaptionskip{-3pt}
\setlength\belowcaptionskip{-5pt}
\caption{ $Q^2$-dependence of the proton and neutron neutral weak Dirac form factor $F^{Z,p(n)}_2$. Statistical and systematic uncertainties are added in quadrature to obtain the final uncertainty. \label{fig:8} }
\end{figure}

We now deduce neutral weak Dirac and Pauli FFs from the above calculation of neutral weak Sachs EMFFs using Eq.~(\ref{SachsFF}). The results are shown in Figures~\ref{fig:figureF1zpn} and~\ref{fig:8}. Similar to the signal-to-noise ratio of the $G^{Z,p}_E(Q^2)$, the signal-to-noise ratio for the proton neutral weak Dirac form factor $F^{Z,p}_1(Q^2)$ decreases with $Q^2$ and the precision is about $3\sigma$ at $Q^2=0.5\,\text{GeV}^2$ after the systematic uncertainties are added in quadrature with the statistical uncertainties. It is shown in Figures~\ref{fig:3} and~\ref{fig:4} that the systematic uncertainties from the LFHQCD dominate over the statistical uncertainties coming from the lattice QCD analysis. Therefore we choose to add the systematic and statistical uncertainties in quadrature to obtain the final error estimates of neutral weak FFs at each $Q^2$-value and present total uncertainties in Figures~\ref{fig:figureF1zpn} and~\ref{fig:8}. These values of the neutral weak FFs are yet to be compared with future experimental determinations since currently these values are experimentally unknown. 

\section{Conclusions}
This analysis presents the determination of the $Q^2$-dependence of the neutral weak electromagnetic form factors. The nucleon neutral weak form factors have been calculated in the momentum transfer range of $0\leq Q^2 \leq 0.5 \,\text{GeV}^2 $ by combining results from light-front holographic QCD and lattice QCD calculations. We have presented a first-principles determination of the $Q^2$-dependence of the strange quark form factors at the physical pion mass $m_\pi=140$ MeV and in the continuum limit. With a model-independent extraction of the $Q^2$-dependence of the strange quark form factors, 24-valence quark masses including at the physical pion mass have been used to explore the quark mass dependence and with finite lattice spacing and volume corrections to determine the strange quark form factors from lattice QCD calculation. Since the strange quark contribution to nucleon electromagnetic form factors are constrained to be small by the global experimental data, and a similar small contribution has been confirmed by first-principles lattice QCD calculations, a precise experimental determination of the neutral weak form factors from parity violating experiments will be challenging. The lattice results of the strange quark form factors at the physical point and light-front holographic QCD prediction of the nucleon electromagnetic form factors have been used to determine the nucleon neutral weak form factors precisely. The determination of neutral weak form factors in this way does not require a prior knowledge of the weak axial form factor and its higher order radiative corrections which are less accurately constrained. Given our precise predictions for the neutral weak electromagnetic form factors at the physical point, we anticipate these results will be verified by future precision experiments.\\ \\


{\bf ACKNOWLEDGMENTS}
R.S.S. thanks the RBC/UKQCD collaborations for providing their DWF gauge configurations. R.S.S. also thanks Stanley J Brodsky, John Connell, Sumit R. Das, Alexandre Deur, Hans G\"{u}nter Dosch, Terrence Draper, Hasnain Hafiz, Mohammad Tariqul Islam, Keh-Fei Liu, Tianbo Liu, Sumiran Pujari, Omer Rahman, and Guy F. de T\'eramond who provided insight and expertise that greatly assisted
the research. This work is supported in part by the U.S. DOE Grant No. DE-SC0013065. This research used resources of the Oak Ridge Leadership Computing Facility at the Oak Ridge National Laboratory, which is supported by the Office of Science of the U.S. Department of Energy under Contract No. DE-AC05-00OR22725. This work used Stampede time under the Extreme Science and Engineering Discovery Environment (XSEDE), which is supported by National Science Foundation grant number ACI-1053575. We also thank National Energy Research Scientific Computing Center (NERSC) for providing HPC resources that have contributed to the research results reported within this paper. We acknowledge the facilities of the USQCD Collaboration used for this research in part, which are funded by the Office of Science of the U.S. Department of Energy.

\bibliographystyle{h-physrev3}

\bibliography{2014_strange_charm_quark_spin}

\end{document}